\documentclass[twocolumn,superscriptaddress,showpacs,prd,aps,amsmath,amssymb,nofootinbib]{revtex4-1}

\usepackage{graphicx}
\usepackage{longtable}

\newcommand{\beq}{\begin{equation}} 
\newcommand{\eeq}{\end{equation}} 
\newcommand{\beqn}{\begin{eqnarray}} 
\newcommand{\eeqn}{\end{eqnarray}}

\newcommand{\zD}{{\raise1.0ex\hbox{${}^{\ \circ}$}}\!\!\!\!\!D}
\newcommand{\alone}{{\raise0.5ex\hbox{${}^{\ 1}$}}\!\!\!\!\alpha}

\newcommand{\nalam}{\mathrel{\raise0.9ex\hbox{$^\lambda$}\mkern-14mu
\lower0.0ex\hbox{$\nabla$}}}

\newcommand{\zeroD}{{\raise1.0ex\hbox{${}^{\ \circ}$}}\!\!\!\!\!D}
\newcommand{\zLap}{{\raise1.0ex\hbox{${}^{\ \circ}$}}\!\!\!\!\Delta}
\newcommand{\zna}{{\raise1.0ex\hbox{${}^{\ \circ}$}}\!\!\!\!\!\nabla}
\newcommand{\zS}{{\raise1.0ex\hbox{${}^{\ \circ}$}}\!\!\!\!\!S}

\usepackage[normalem]{ulem}
\usepackage{epstopdf}
\usepackage{color}
\usepackage{soul}
\usepackage{ulem}
\usepackage{bm}
\usepackage{xspace}

\newcommand{\SACRA}{\textsc{sacra-mpi}\xspace}
\newcommand{\cocal}{\textsc{cocal}\xspace}

\usepackage[]{amsmath}
\usepackage[]{amsfonts}
\usepackage[]{amssymb}
\usepackage[]{mathrsfs}
\usepackage[]{mathtools}
\usepackage{times}
\usepackage{hyperref}
\hypersetup{
  colorlinks=true,        
  linkcolor=blue,         
  citecolor=cyan,         
}
\newcommand{\apjl}{Astrophys. J. Lett.}

\def\QEQ{{%
			\setbox0\hbox{$I$}%
			\rlap{\hbox to \wd0{\hss--\hss}}\box0
		}}

\begin{document}

\title{Evolution of equal mass binary bare quark stars in full general relativity: could a supramassive merger remnant experience prompt collapse?}

\author{Enping Zhou}
\affiliation{Huazhong University of Science and Technology, School of Physics, 1037 Luoyu Road, Wuhan, 430074, China}
\affiliation{Max Planck Institute for Gravitational Physics (Albert Einstein
Institute), Am M\"uhlenberg 1, Potsdam-Golm, 14476, Germany} 

\author{Kenta Kiuchi}
\affiliation{Max Planck Institute for Gravitational Physics (Albert Einstein
	Institute), Am M\"uhlenberg 1, Potsdam-Golm, 14476, Germany} 
\affiliation{Center
	for Gravitational Physics, Yukawa Institute for Theoretical Physics, Kyoto
	University, Kyoto, 606-8502, Japan}
	
\author{Masaru Shibata}
\affiliation{Max Planck Institute for Gravitational Physics (Albert Einstein
	Institute), Am M\"uhlenberg 1, Potsdam-Golm, 14476, Germany} \affiliation{Center
	for Gravitational Physics, Yukawa Institute for Theoretical Physics, Kyoto
	University, Kyoto, 606-8502, Japan}

\author{Antonios Tsokaros}
\affiliation{Department of Physics, University of Illinois at Urbana-Champaign,
	Urbana, IL 61801, USA}

\author{K\=oji Ury\=u}
\affiliation{Department of Physics, University of the Ryukyus, Senbaru,
	Nishihara, Okinawa 903-0213, Japan}

\date{\today}

\begin{abstract}

We have evolved mergers of equal-mass binary quark stars, the total mass of which is close to the mass shedding limit of uniformly rotating configurations, in fully general relativistic hydrodynamic simulations, aimed at investigating the post-merger outcomes. In particular, we have identified the threshold mass for prompt black hole formation after the merger, by tracing the minimum lapse function as well as the amount of ejected material during the merger simulation. A semi-analytical investigation based on the angular momentum contained in the merger remnant is also performed to verify the results. For the equation of state considered in this work, the maximum mass of TOV solutions for which is 2.10\,$M_\odot$, the threshold mass is found between 3.05 and 3.10\,$M_\odot$. This result is consistent (with a quantitative error smaller than 1\%) with the universal relation derived from the numerical results of symmetric binary neutron star mergers. Contrary to the neutron star case, the threshold mass is close to the mass shedding limit of uniformly rotating quark star. Consequently, we have found that binary quark stars with total mass corresponding to the long-lived supramassive remnant for neutron star case, could experience collapse to black hole within several times dynamical timescale, making quark stars as exceptions of the commonly accepted post-merger scenarios for binary neutron star mergers. We have suggested explanation for both the similarity and the difference, between quark stars and neutron stars.

\end{abstract}

\maketitle

\section{Introduction}
The outcome of a binary neutron star (BNS) merger event could provide rich information on the equation of state (EOS) of cold dense matter 
because the lifetime of the merger remnant before collapsing to black hole (BH) depends on the total mass of the binary system and the EOS of the merging NSs (for recent reviews, see e.g. \citep{Shibata2019rev,Friedman2020,Radice2020,Bernuzzi2020}) 
and the amount and properties of the ejecta and the evolution of the magnetic field during the post-merger stage are closely related to the evolution of the massive remnant, which could be implied from the observations of the electromagnetic counterparts of the BNS merger \citep{Metzger2014,Metzger2014b,Ruiz2017,Shibata2017c,Kiuchi2017}. According to the post-merger outcome as inferred from observations for the BNS merger event GW170817 \citep{Abbott2017,Abbott2017b}, various constraints has been put on the EOS of the merging NSs \citep{Ruiz2017,Rezzolla2017,Bauswein2017b,Margalit2017,Shibata2019,Kiuchi2019}. 

There are several critical masses which determine the fate of the post-merger remnant \citep{Margalit2019}. One is the threshold mass of prompt collapse ($M_\mathrm{thres}$): an immediate formation of BH  will occur if the total mass of the binary ($M_\mathrm{tot}$) is beyond this mass, otherwise a massive NS remnant is formed \citep{Shibata05d,Shibata06a,Baiotti08,Hotokezaka2011,Bauswein2013}. Various attempts have been done in deriving an EOS-insensitive relation for $M_\mathrm{thres}$ so as to make better use of the current and future observations of BNS mergers to constrain the EOS models of NSs \citep{Bauswein2013,Bauswein2017,Bauswein2020,Bauswein2020b,Koeppel2019}. In the delayed collapse case, if the remnant could be supported by only uniform rotation (i.e., the mass of the remnant is less than the mass shedding limit of uniform rotation $M_\mathrm{max,urot}$) \citep{Cook94c,Lasota1996}, it is called a supramssive NS (SMNS). Otherwise it is called a hypermassive NS (HMNS) and has to be supported by differential rotation \citep{Baumgarte00bb,Kastaun2014,Hanauske2016}. The angular momentum of the merger remnant is expected to be redistributed quickly by effective turbulent viscosity produced by magnetohydrodynamical instabilities and gravitational torque. The remnant is expected to be back to uniform rotation in a timescale of $\sim$100\,ms \citep{Shibata2017d,Fujibayashi2017b}. Although SMNS could further dissipate its angular momentum due to magnetic braking, the lifetime of a SMNS remnant should be much longer than HMNS case before undergoing a delayed collapse to BH. There is another possibility of a stable NS remnant if the remnant mass is smaller than the maximum mass of cold spherical NS configurations (i.e., $M_\mathrm{max,TOV}$).

In addition to conventional NS models, many efforts have been made in the role of a strong interaction phase transition (PT), which might lead to the existence of a deconfined quark phase \citep{Annala2020}, on the post-merger outcome \citep{Bauswein2019b,Most2019}. Nevertheless, the formation of a deconfined quark core in the high density part of the remnant is not the only possible outcome of the strong interaction PT, the possibility of a self-bound compact star composed entirely of deconfined $u,d,s$ quarks (i.e., bare quark star) has been suggested \citep{Bodmer1971,Witten84,Alcock86,Drago2016c,Drago2016d,Bhattacharyya2017}. However, due to the self-bound nature of such objects, there exists a density discontinuity on the surface of quark stars (QS) which has made it difficult to be evolved with general relativistic hydrodynamic (GRHD) simulations. In spite of very few previous attempts in simulating binary merger involving QSs (cf. \citep{Bauswein2010} in approximate gravity and smooth particle hydrodynamics), the dynamics of BQS mergers is still poorly understood. 

In particular, it is known that by uniform rotation, the maximum mass of QSs 
is enhanced much more than NSs (i.e., $M_\mathrm{max,urot}/M_\mathrm{max,TOV}$ is typically 1.2 for NSs, but 1.4 for QSs \citep{Li2016}) and $M_\mathrm{max,urot}$ for QSs could be very close to $M_\mathrm{thres}$ supposing that QSs follow the same universal relation of $M_\mathrm{thres}$ for prompt collapse  (which predicts $M_\mathrm{thres}/M_\mathrm{max,TOV}\approx 1.46$ for the QS model considered in this paper). This indicates that either QSs do not follow the universal relation of threshold mass for prompt collapse as NSs, or the scenario of post-merger outcome is totally different from NSs, as in BNS cases the gap between $M_\mathrm{thres}$ ($M_\mathrm{thres}/M_\mathrm{max,TOV}\approx 1.3-1.7$ \citep{Hotokezaka2011}) and $M_\mathrm{max,urot}$ is much larger than QS case and a system with a total mass in between those two values corresponds to the HMNS remnant scenario. It is suggested that mergers resulting in prompt collapse or a hypermassive remnant will lead to very distinguished observational features of the short gamma-ray burst and kilonova counterparts \citep{Ruiz2017,Metzger2014,Metzger2014b,Margalit2019}, and a particular caution was made for unequal-mass merger cases \citep{Kiuchi2019}. And hence, verifying the post-merger scenario with appropriate GRHD simulation of BQS mergers with total mass close to $M_\mathrm{max,urot}$ is very important for the interpretation of future observations of binary mergers.

Very recently, we have developed a novel numerical approach for handling bare QSs in numerical relativity simulations \citep{Zhou2021}, which enabled us to explore the dynamics of BQS mergers in grid-based GRHD simulations for the first time. We have employed our methods in the evolution of BQS mergers with different masses, aimed at deriving the threshold mass of prompt collapse in BQS mergers. The EOS and binary models used in this work will be introduced in Sec.~\ref{sec:model}. The simulation results and interpretation of them will be reported in Sec.~\ref{sec:result}. Finally,  conclusions and discussions about the impact of our results on understanding future observations will be described in Sec.~\ref{sec:dc}.

\section{The models}
\label{sec:model}

Similar to our previous work, MIT bag model \citep{chodos1974} is used as the EOS for the merging QSs in our simulations. Modern details of the model, such as finite mass of strange quarks and gluon-mediated interactions \citep{Fraga2001,Alford2005}, are not considered for simplicity. We have extensively tested the reliability and accuracy of handling the simplest form of the MIT bag model with our numerical implementation in our previous study~\citep{Zhou2021}. Hence, we stick with the same model: 
\beq
\begin{split}
	p_\mathrm{cold}=K\rho^{4/3}-B, \\
	e_\mathrm{cold}=3K\rho^{4/3}+B,
\end{split}
\label{eq:mitexplicit}
\eeq
in which $p_\mathrm{cold}$, $\rho$ and $e_\mathrm{cold}$ are the pressure, rest-mass density and energy density of quark matter at zero temperature, respectively. $B$ is the bag constant and is set to be $52.5\,\mathrm{MeV\,fm^{-3}}$. The parameter $K$ is determined as
\beq
K=\left(\frac{c^8}{256B}\right)^{1/3},
\eeq
in order to resolve the discontinuity of specific enthalpy across the surface of the QSs (the details are found in \citep{Zhou2021}). With such a choice, $M_\mathrm{max,TOV}$ is 2.10\,$M_\odot$ and the tidal deformability for a 1.4\,$M_\odot$ star is 598.  To take the finite temperature effects into account, we have applied the $\Gamma_\mathrm{th}$ prescription in which the total pressure is described as a sum of the cold part $p_\mathrm{cold}(\rho)$ of Eq.~(\ref{eq:mitexplicit}) and a thermal part in the form:
\beq
p=p_\mathrm{cold}+(\Gamma_\mathrm{th}-1)\rho\epsilon_\mathrm{th},
\label{gammath}
\eeq
where $\epsilon_\mathrm{th}$ is the thermal part of the specific energy density defined by the difference between the total specific energy density and the cold part as obtained from Eq.~(\ref{eq:mitexplicit}). The index $\Gamma_\mathrm{th}$ is chosen to be 4/3 such that the primitive recovery procedure inside the QS is essentially solving a quadratic equation which allows us to accurately resolve whether or not a fluid element is inside the QS without losing computational efficiency \citep{Zhou2021}. 

The value of $M_\mathrm{max,urot}$ with this EOS is 3.03\,$M_\odot$, and with differential rotation the mass of QSs could be much higher. Suppose that BQS merger follows the same universal relation of $M_\mathrm{thres}$ as BNS mergers (e.g. as in \citep{Bauswein2017}), the derived value of  $M_\mathrm{thres}$ for this model would be approximately 3.07\,$M_\odot$, i.e., very close to $M_\mathrm{max,urot}$. It is therefore interesting to verify what happens for the remnant of a BQS merger with a total mass near the range of $M_\mathrm{max,urot}$ and $M_\mathrm{thres}$, and to figure out a more accurate value of $M_\mathrm{thres}$ 
by fully GRHD simulations. 

For this purpose, we have prepared irrotational equal-mass BQS initial data with different total masses (3.0\,$M_\odot$, 3.05\,$M_\odot$ and 3.1\,$M_\odot$ at infinite separation, and the models are referred to as MIT3.00, MIT3.05 and MIT3.10 through the texts) separated at an initial distance of 35\,km with the \cocal code \citep{Tsokaros2015,Uryu2012}. The surface fit coordinate employed in \cocal code is essential for obtaining accurate initial data for QSs which have a finite surface density \citep{Zhou2018,Zhou2019}. The convergence and accuracy of \cocal code in calculating initial data for co-rotational, irrotational and spinning BQSs will be reported in another paper \citep{Tsokaros2021}. The evolution is done with the \SACRA code \citep{Yamamoto2008,Kiuchi2017b}, which employs a moving puncture version of the Baumgarte-Shapiro-Shibata-Nakamura formalism~\citep{Shibata95,Baumgarte99,Campanelli06,Baker:2005vv} and a constraint propagation prescription similar to the Z4c scheme \citep{Hilditch2012} to solve the Einstein's evolution equations. In order to evolve bare QSs with finite surface density, an essential modification in the primitive recovery part of the code is done and reported in our previous research~\citep{Zhou2021}. An adaptive mesh refined setup is implemented for the evolution of the BQS models, the details of which are found in~\citep{Yamamoto2008}. For the highest resolution runs, 
the resolution and size of the finest level are $\Delta x=\Delta y=\Delta z=74\,$m and $L_x=L_y=2L_z=23.68\,$km (orbital plane symmetry is assumed for the simulations). For comparison, we have done all the simulations in another grid setup with $\Delta x=\Delta y=\Delta z=118.4\,$m.  

To verify whether or not a prompt collapse happens for a certain model, we have tracked the minimum lapse function $\alpha_\mathrm{min}$ throughout the simulation. It is a robust indicator for the appearance of an apparent horizon \citep{Alcubierre:2008}. In addition, for equal-mass binaries, the bounce of $\alpha_\mathrm{min}$ is related to the bounce of the remnant in the post-merger, which would drive a significant fraction of mass ejection during the merger. By contrast, in the prompt collapse case, the bounce is absent and the remnant straightforwardly collapses to a BH. Thus the mass ejection is negligible for equal mass mergers. According to this, we have tried to identify the prompt collapse by searching for the model with least mass when no bounce in $\alpha_\mathrm{min}$ is observed.

\section{Results}
\label{sec:result}

\begin{figure}
	\begin{center}
		\includegraphics[height=70mm]{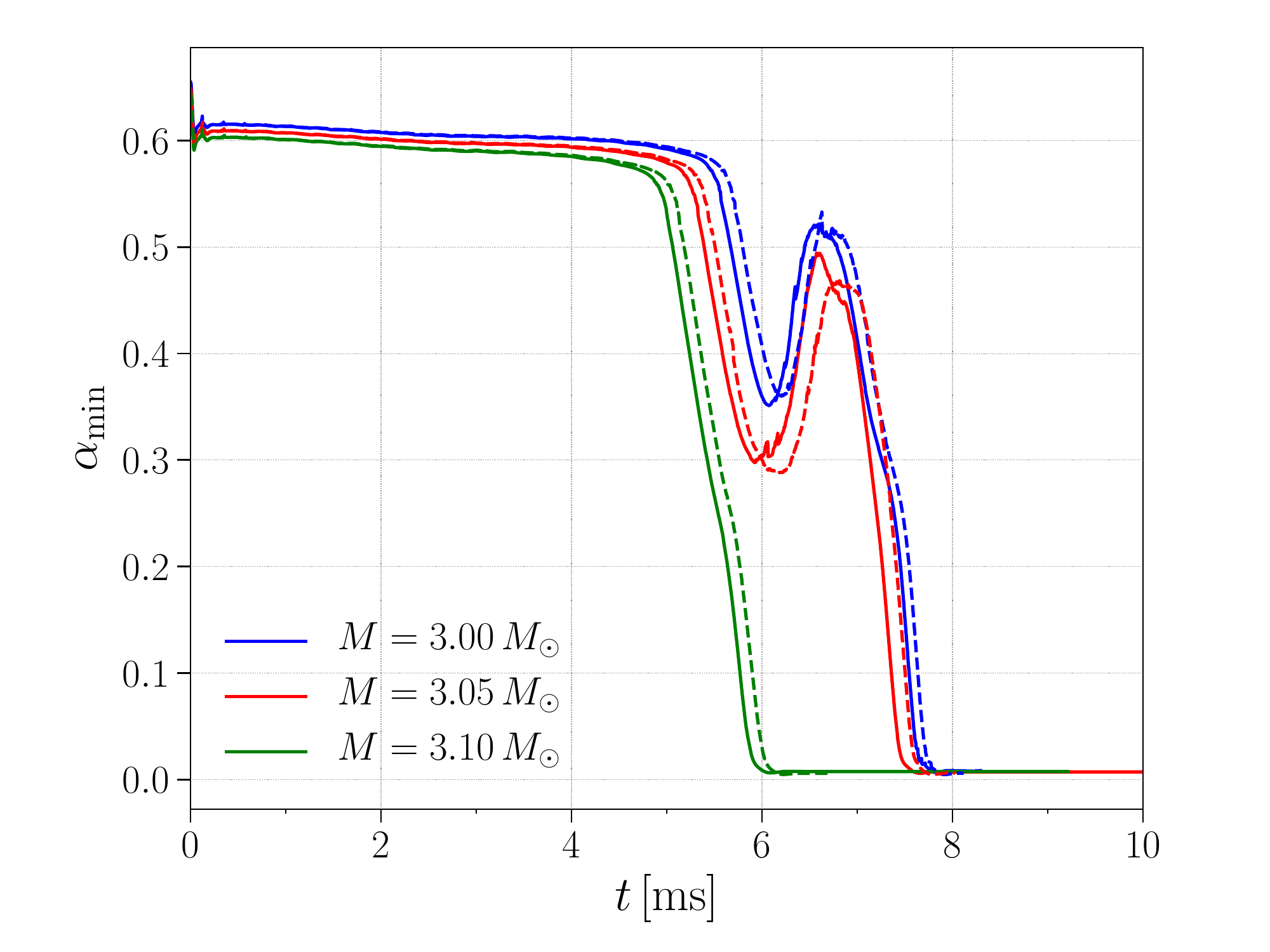}
	\end{center}
	\caption{The evolution of the minimum lapse function for the three models  MIT3.00 (blue curve), MIT3.05 (red curve) and MIT3.10 (green curve). The results in the low and high resolution are shown in the dashed and solid curves, respectively.}
	\label{fig:alphamin}
\end{figure}

Figure~\ref{fig:alphamin} shows the evolution of $\alpha_\mathrm{min}$ for the 3 models with different total mass and in two different resolutions. As can be seen, there is one bounce for $\alpha_\mathrm{min}$ before the collapse to BH for the models with total mass of 3.0 and 3.05\,$M_\odot$. By contrast, for the model with 3.1\,$M_\odot$, the merger remnant collapses
to a BH without any bounce, constraining $M_\mathrm{thres}$ for this EOS model to be between 3.05 and 3.10\,$M_\odot$. This result is universal in the two different resolutions we considered.

The lifetime of the merger remnant before collapsing to BH could be indicated by the duration between the first local minimum of $\alpha_\mathrm{min}$ and the moment it approaches zero. According to Fig.~\ref{fig:alphamin}, it is found that the lifetime of the remnant for the models MIT3.00 and MIT3.05 are similar, i.e., approximately 1\,ms, and all correspond to a very short-lived case. On the other hand, the lifetime of the remnant for MIT3.10 vanishes, which is the consistent with a prompt collapse scenario.

\begin{figure}
	\begin{center}
		\includegraphics[height=70mm]{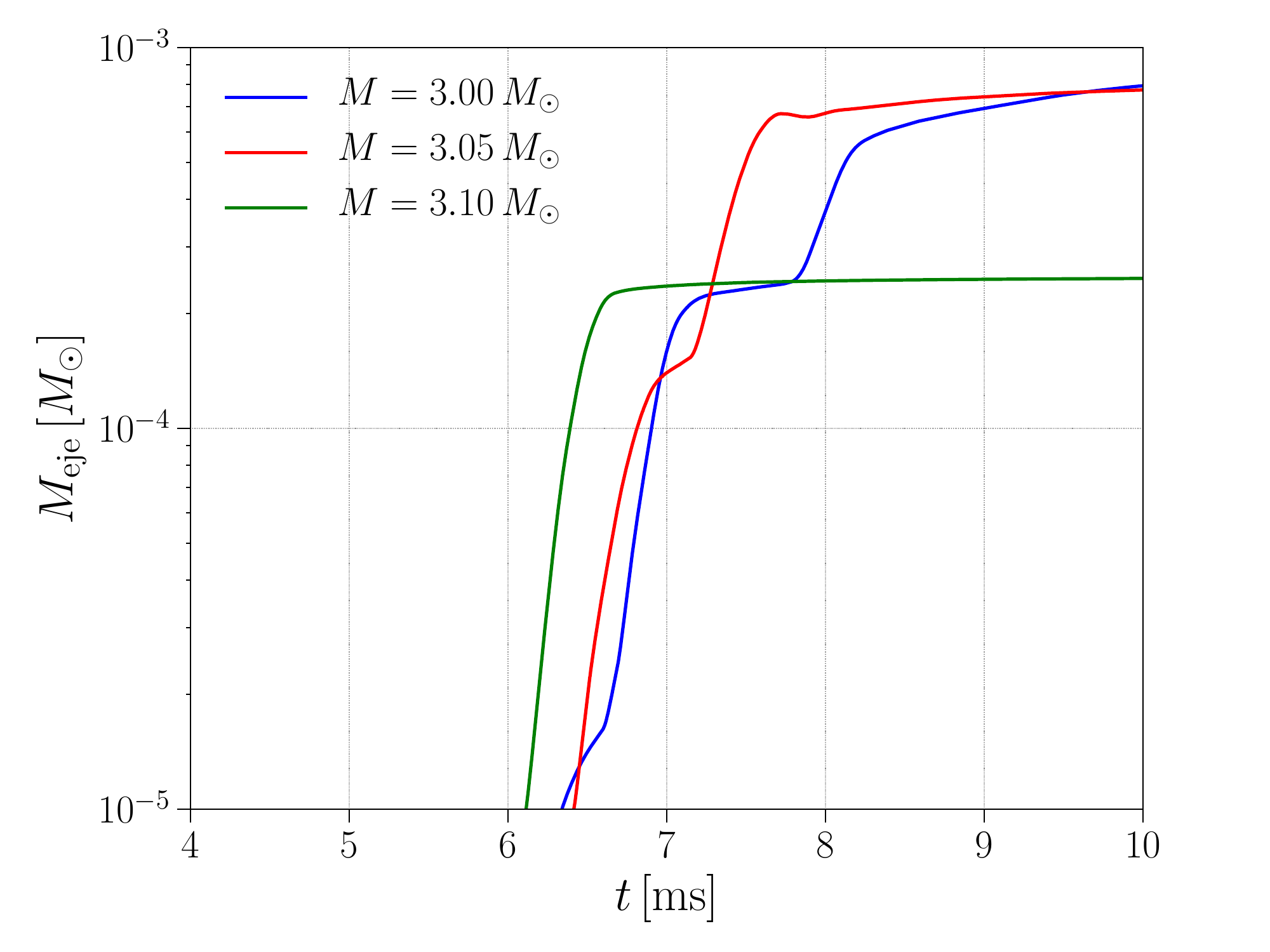}
	\end{center}
	\caption{The amount of unbound material (as estimated with the criterion that $u_t<-1,0$ and distance from the coordinate origin $d>147.7\,$km) for the three models considered in this work, for the high resolution runs.}
	\label{fig:ejecta}
\end{figure}

In addition to the evolution of $\alpha_\mathrm{min}$, we also display the time evolution of 
the ejecta mass of each model in Fig.~\ref{fig:ejecta}. This shows that the 
ejecta mass is tightly related to the post-merger outcome. Here, the ejecta mass 
is estimated by integrating all matter with $u_t < -1.0$ and distance larger than $147.7\,$km from the coordinate origin. As can be seen from the figure, for model MIT3.10, the rising of the amount of ejecta happens earlier as the merger happens at an earlier time compared with the other two models (cf. the moment when $\alpha_\mathrm{min}$ starts to rapidly drop in Fig.~\ref{fig:alphamin}) but saturates at a lower total amount when the BH is eventually formed due to the absence of the bounce in the merger remnant. Similar to what is seen in the lifetime of the merger remnant, the ejecta mass is nearly the same for model MIT3.00 and MIT3.05, while the difference between the MIT3.10 and the other two models are much larger than the difference between MIT3.00 and MIT3.05. These features all indicate a prompt collapse for the MIT3.10 case. 

\begin{figure}
	\begin{center}
		\includegraphics[height=70mm]{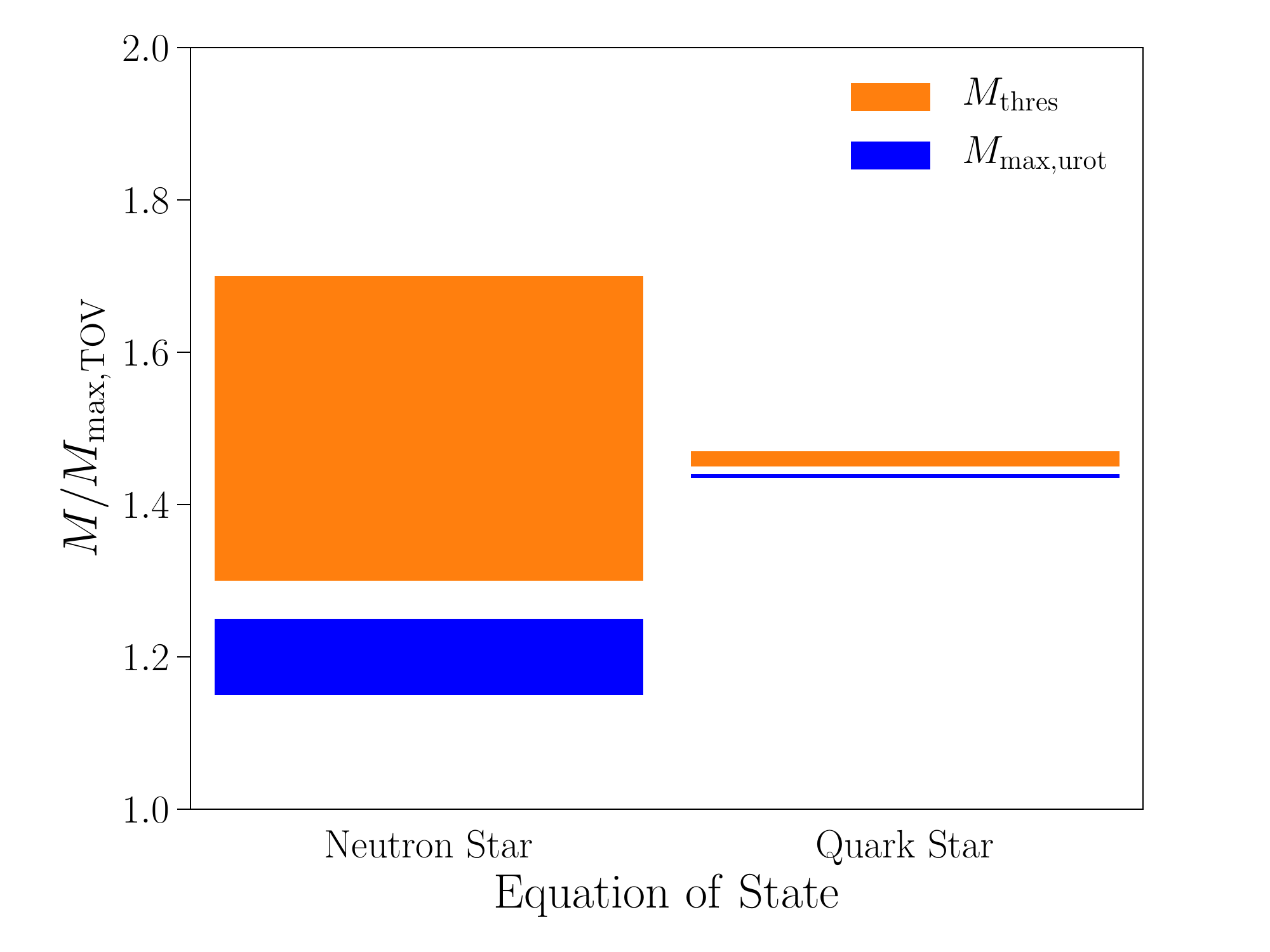}
	\end{center}
	\caption{A quantitative comparison between $M_\mathrm{thres}$ (orange shaded area) and $M_\mathrm{max,urot}$ (blue shaded area) of NSs (shown on the left half) and QSs (on the right half). The plots for QSs are made according to the results obtained in this work for the MIT bag model and those for NSs are according to previous researches on various NS EOS models (and hence being more scattered) \citep{Hotokezaka2011,Breu2016}. It is worth noting that, for any particular NS EOS model, $M_\mathrm{thres}$ and $M_\mathrm{max,urot}$ take their unique value from the shaded range in the figure. Therefore, the gap between the two shaded areas for the NS case is merely the lower limit for the difference between $M_\mathrm{thres}$ and $M_\mathrm{max,urot}$ for any given EOS model, yet still being much larger than the gap of the QS model considered in this work.}
	\label{fig:illustration}
\end{figure}

According to what is discussed above, the threshold mass for prompt collapse of the MIT bag model we are considering is between 3.05 and 3.10\,$M_\odot$. In another word, a BQS whose total mass is very close to the maximum mass of uniformly rotating configuration experiences prompt collapse in the post-merger phase. As shown in Fig.~\ref{fig:illustration}, it is quite different from BNS cases, for which $M_\mathrm{thres}$ is much larger than $M_\mathrm{max,urot}$ and a BNS merger with mass close to $M_\mathrm{max,urot}$ is expected to produce a relatively long-lived remnant which can survive after the dissipation of the differential rotation. In fact, model MIT3.00, which by definition corresponds to the supramassive remnant scenario, collapses to BH within 2\,ms, becoming incompatible with the post-merger scenarios of BNS. On the contrary, quantitatively the threshold mass is in good agreement with the universal relation derived from BNS simulations. To understand the difference and similarity of BQS and BNS mergers, we have analyzed the mass and angular momentum in the merger remnant and come up with a semi-analytical explanation, 
in the similar approach to \citep{Bauswein2017}.

\begin{figure}
	\begin{center}
		\includegraphics[height=70mm]{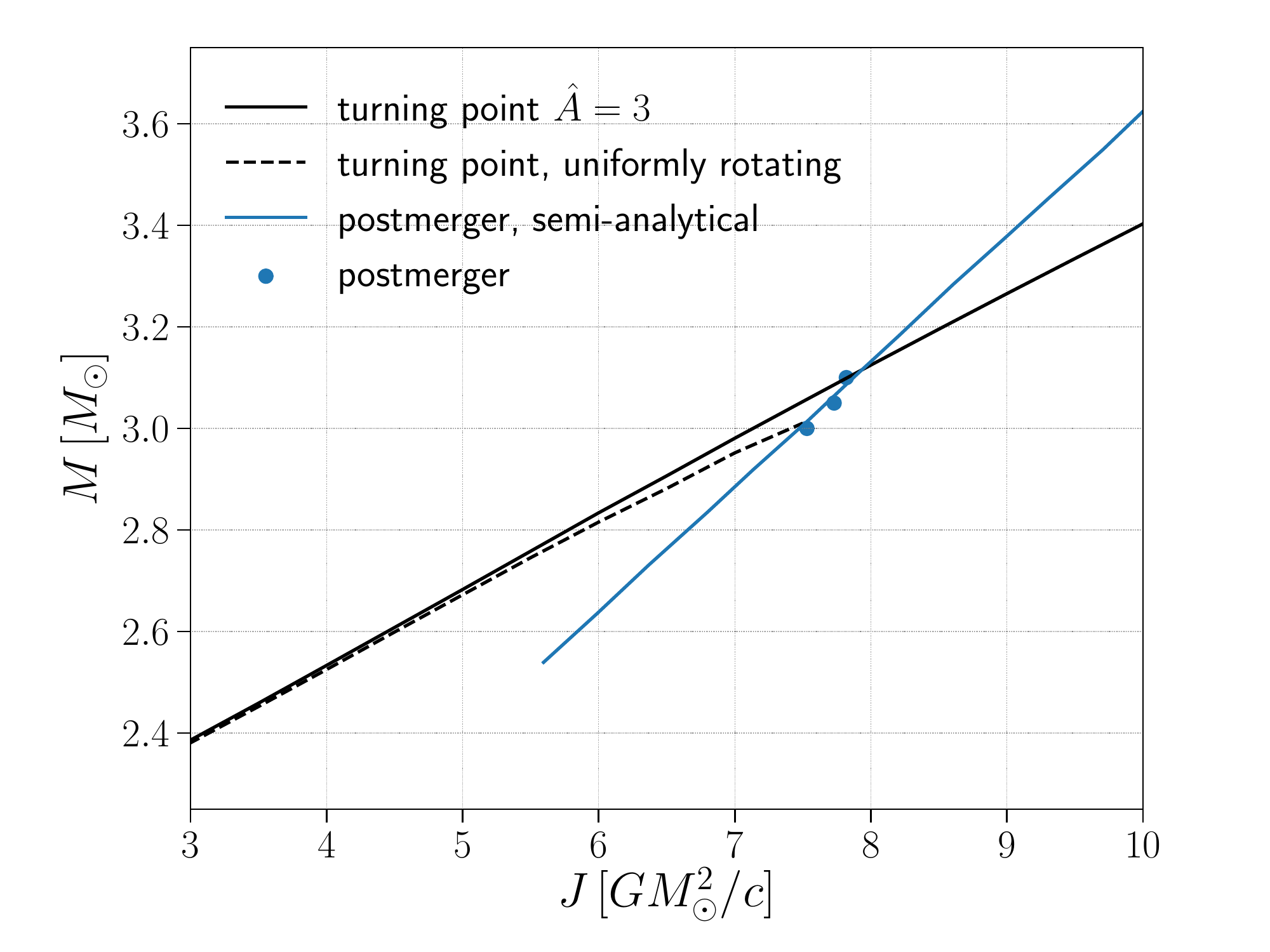}
	\end{center}
	\caption{A semi-analytical approach to understand the post-merger outcome of BQS mergers. The solid/dashed black curves are the angular momentum and mass for the "turning point" solutions of differentially/uniformly rotating QSs. $j-$const is chosen as the differential rotation law with $\hat{A}=3.0$. The blue curve is the empirical relation of the angular momentum and mass for merger remnant based on simulations of BNS mergers, according to previous researches \citep{Bauswein2017}. The intersection of the blue curve and "turning point" curve should indicate the threshold mass, as beyond this mass, the merger remnant would not be able to have sufficient angular momentum to support itself. The blue filled circles are the angular momentum and mass for the merger remnant as extracted from the BQS simulations considered in this work.}
	\label{fig:semiana}
\end{figure}

Figure~\ref{fig:semiana} shows the angular momentum and mass of the merger remnant (estimated by subtracting out the angular momentum and energy radiated by GW during the inspiral stage from the initial values) with filled circles, at the moment when merger happens (by choosing the time when the gravitational wave strain reaches maximum at the end of inspiral stage), for the 3 models considered in this work. In addition, we plot the empirical relation between the angular momentum and mass of the merger remnant for BNS, fitted by the simulation results with various NS EoS \citep{Bauswein2017}. As immediately found, the results of BQS merger simulations are in a good agreement with the BNS empirical relation, which is a key
for understanding the fact that $M_\mathrm{thres}$ of BQS mergers follows the universal relation for BNS cases. The intersection of the blue curve and the turning point solutions of differentially/uniformly (indicated black solid/dashed line) rotating QS should approximately indicate the threshold mass of BQS cases, as beyond the intersection point, the merger remnant would not have enough angular momentum to support the mass it possesses. Indeed, the intersection happens between 3.0 and 3.1 solar mass, which again confirms our fully GRHD simulation results.

Combining the simulation results and the semi-analytical explanation, we are now able to understand the post-merger scenario of BQS mergers. Although QSs could reach much higher maximum mass with uniform/differential rotation compared with NSs, these equilibrium configurations do need much higher angular momentum to support the mass \citep{Zhou2019}. Nevertheless, for a fixed total mass of the binary, the final angular momentum contained in the merger remnant is similar for BQS and BNS cases. Hence, a BQS system with mass close to $M_\mathrm{max,urot}$ could experience prompt collapse since the merger remnant would not be able to have angular momentum sufficient for supporting itself. More specifically, the angular momentum contained in the merger remnant is related to how much angular momentum is radiated by GW during the inspiral, which could be well modeled by the masses of the binary components and finite size effects, such as the tidal deformability. Therefore, it is not very surprising that $M_\mathrm{thres}$ of BQS mergers could also be estimated with the universal relation for BNS mergers, as tidal deformability/compactness is needed for the universal relation.

In addition, it is worth mentioning that quantities of individual QS as described by Eq.~(\ref{eq:mitexplicit}) scales with the bag constant $B$ for both non-rotating and rotating configurations \citep{Bhattacharyya2016}, in particular, mass and radius of a solution scales as
\beq
(M_1, R_1) =\sqrt{B_2/B_1}(M_2, R_2)
\eeq
for solutions with scaled central density and angular velocity:
\beq
\begin{split}
\rho_\mathrm{c,1}=(B_1/B_2)\rho_\mathrm{c,2}, \\
\Omega_1=\sqrt{B_1/B_2}\Omega_2.
\end{split}
\eeq
Nevertheless, the result we obtained in this work (i.e., $M_\mathrm{thres}$ of BQS mergers follows the universal relation derived from BNS simulations, and is very close to $M_\mathrm{max,urot}$) is not a coincidence due to a particular choice of $B$, as a similar quantitative result was obtained in \citep{Bauswein2010} for larger choices of $B$. This could help us verify the applicability of the universal relation of $M_\mathrm{thres}$ for QSs modeled with different bag constants. On the one hand, due to the same scaling of both non-rotating and rotating configurations, the ratio $M_\mathrm{max,urot}/M_\mathrm{max,TOV}$ is independent of $B$. On the other hand, the ratio between $M_\mathrm{thres}$ and $M_\mathrm{max,TOV}$ could be obtained by universal relation $k=M_\mathrm{thres}/M_\mathrm{max,TOV}$. In the cases that $k$ is fitted with the compactness/tidal deformability of a TOV maximum mass solution, it is also very insensitive to the choice of $B$, leading to a similar value of $M_\mathrm{thres}$ as $M_\mathrm{max,urot}$ for all choices of bag constant.

\section{Conclusions and Discussions}
\label{sec:dc}

In this work, we performed fully GRHD simulations of BQS mergers with different masses and tried to determine the threshold mass of prompt BH formation of BQS mergers, which is a key parameter for constraining EoS of compact stars with future multi-messenger observations, by investigating quantities such as the minimum lapse function and ejecta amount during the post-merger stage. The value of $M_\mathrm{thres}$ is found to be between 3.05 and $3.10M_\odot$ with respect to $M_{\rm max,TOV} \approx 2.10M_\odot$
for the EoS model we considered, and is in good agreement with the universal relation derived with various BNS simulations \citep{Bauswein2017}. This is due to the fact that the inspiral-in of a binary compact star is a purely 'gravity-bound' process: the dynamics during this stage is determined by the gravitational interaction between the two stars, and any information about the EoS and structure of those merging stars, are concentrated in the quantities such as tidal deformability and compactness. Consequently, in spite of the fact that bare QSs are self-bound and could reach more massive rotating configurations by containing more angular momentum, the angular momentum left in the merger remnant is similar to the case of BNS, as long as the total mass and tidal deformability of the merging binary are similar. Due to the same reason, a supramassive remnant of BQS merger could collapse to a BH within a few times the dynamical timescale, as the mass shedding limit of uniformly rotating QSs demands more angular momentum than what could be left in the remnant.

According to our results, for any future multi-messenger observations which we could confirm whether or not a prompt collapse happens, the derived radius/compactness/tidal deformability constraint for NSs according to the merger outcome could be directly applied for QSs as well, as the universal relations of  $M_\mathrm{thres}$ is valid for both cases. 
Nevertheless, as we are now only beginning to attempt simulations of BQS mergers in fully GRHD, there is still a lot to improve in our understanding of what happens during a BQS merger. For instance, although the amount of ejected material is found to be in similar order of magnitude compared with BNS cases, it is not straightforward to predict the signatures of the EM counterparts, as the nuclear physical evolution of those ejected quark matter is quite unclear \citep{Madsen1988,Caldwell1991,Paulucci2017,Bucciantini2019}. More sophisticated nuclear physical models and nucleosynthesis are required to be combined with the simulation results for us to have a better understanding about the EM signatures of BQS mergers. In addition, limited by our current primitive recovery scheme for QSs, it is quite complicated and time-consuming to explore the impact of the thermal components. We will extend our study to more general cases so as to systematically explore the impact of thermal component and mass ratio on the dynamics of BQS mergers in the future.

\acknowledgements

We thank the members of Computational Relativistic Astrophysics division in Max Planck Institute for Gravitational Physics (Potsdam) for very helpful discussions. E. Z. acknowledges the support by National SKA Program of China No. 2020SKA0120300. This work is also in part supported
by Grant-in-Aid for Scientific Research (Grant No. 20H00158,
Nos.JP18H01213, 18K03624, and 20H00158) of Japanese MEXT/JSPS. A.T. is supported in part by National Science Foundation (NSF) Grant
PHY-1662211 and PHY-2006066, and NASA Grant 80NSSC17K0070 to
the University of Illinois at Urbana-Champaign. The simulations are performed on the SAKURA at Max Planck Computing and Data Facility (MPCDF). 

\bibliographystyle{apsrev4-1}
\bibliography{aeireferences}

\end{document}